\documentstyle[12pt,epsf,epsfig,wrapfig]{article}

\begin{document}

\thispagestyle{empty}
\renewcommand{\thefootnote}{\fnsymbol{footnote}}
\begin{flushleft}
\large {SAGA-HE-109-96  \hfill 
October 27, 1996} \\
\end{flushleft}
 
\vspace{2.0cm}
 
\begin{center}
{\Large \bf Numerical solution of NLO $\bf Q^2$ evolution equations \\ }

\vspace{0.2cm}
{\Large \bf for spin-dependent structure functions \\ }

\vspace{1.8cm}

\Large
M. Hirai, S. Kumano, and M. Miyama $^*$

\vspace{1.0cm}
 
Department of Physics  \\
 
\vspace{0.1cm}
 
Saga University \\
 
\vspace{0.1cm}

Saga 840, Japan \\

\vspace{1.5cm}

\large 
Talk given at the 12th International Symposium on \\

\vspace{0.3cm}

High-Energy Spin Physics \\

\vspace{0.7cm}

Amsterdam, The Netherlands, September 10 $-$ 14, 1996 \\
(talk on Sept. 12, 1996) \\
 
\end{center}

\vspace{1.5cm}

\vfill
 
\noindent
{\rule{6.cm}{0.1mm}} \\
 
\vspace{-0.2cm}
\normalsize
\noindent
{* Email: 96sm18, kumanos, 96td25@cc.saga-u.ac.jp.} \\

\vspace{-0.6cm}
{Information on their research is available at}  \\

\vspace{-0.6cm}
{http://www.cc.saga-u.ac.jp/saga-u/riko/physics/quantum1/structure.html.} \\

\vspace{+0.5cm}
\hfill
{to be published in proceedings}

 \clearpage
\setcounter{page}{1}

\begin{center}
{\large \bf Numerical solution of NLO $\bf Q^2$ evolution equations \\ }
{\large \bf for spin-dependent structure functions \\ }
\vspace{5mm}

M. Hirai, S. Kumano, and M. Miyama $^*$

\vspace{5mm}
{\small\it
Department of Physics, Saga University, Saga 840, Japan
\\ }

\vspace{5mm}
ABSTRACT

\vspace{5mm}
\begin{minipage}{130 mm}
\small


Numerical solution of DGLAP $Q^2$ evolution equations
is studied for polarized parton distributions by using
a ``brute-force" method.
NLO contributions to splitting functions are recently calculated,
and they are included in our analysis.
Numerical results in polarized parton distributions and
in the structure function $g_1$ are shown.
In particular, we discuss how numerical accuracy
depends on number of steps in the variable $x$ and in $Q^2$. 
 
\end{minipage}
\end{center}

Spin-dependent structure functions are measured in
polarized lepton-nucleon scattering. 
They depend on two kinematical variables,
$x$ and $Q^2$. The $Q^2$ dependence is calculated within
perturbative QCD, and it is described by
integrodifferential equations so called DGLAP equations.
Because the $Q^2$ evolution equations are often used in
theoretical and experimental studies, it is useful to have
a computer code for solving the equations numerically.
Our studies are important, for example, in getting optimal
parton distributions by analyzing $g_1$ experimental data,
particularly in obtaining gluon polarization by studying scaling
violation of $g_1$.

In general, the DGLAP equations are coupled integrodifferential equations:
$$
{\partial \over {\partial t}} \,  \Delta {q}_i \left({x,t}\right)\, =\,  
\int_{x}^{1}{dy \over y}\,  
\left[{\, \sum_j  {\Delta P}_{q_{i} q_{j}}\left({{x \over y}}\right)\,  
 \Delta q_j \left({y,t}\right)\,  
+\,    {\Delta P}_{qg}\left({{x \over y}}\right)\,  
 \Delta g\left({y,t}\right)\,  }\right]
\ \ ,
\eqno{(1a)}
$$
$$
{\partial \over {\partial t}} \, \Delta g\left({x,t}\right)\, =\,  
\int_{x}^{1}{dy \over y}\,  
\left[{\,  \sum_j {\Delta P}_{gq_j}\left({{x \over y}}\right)\,  
 \Delta q_j\left({y,t}\right)\,  
+\,  {\Delta P}_{gg}\left({{x \over y}}\right)\, 
 \Delta g\left({y,t}\right)\,  }\right]
\ \ .
\eqno{(1b)}
$$
where the variable $t$ is defined by
$t = -(2/ \beta_0)\ln [\alpha_s(Q^2)/\alpha_s(Q_0^2)]$,
$\Delta q_j (x,t)$ and $\Delta g(x,t)$ are 
polarized $j$-flavor quark and gluon distributions,
and $\Delta P_{ij}(x)$ are splitting functions.
Each term in Eqs. (1a) and (1b) describes the process
that a parton $p_j$ with the nucleon's momentum fraction $y$ 
splits into a parton $p_i$ with the momentum fraction $x$
and another parton.
The splitting function determines the probability of 
such a splitting process. 

Next-to-leading-order (NLO) splitting functions
for spin-dependent parton distributions are evaluated recently [1].
Therefore, we can study numerical solution of the NLO
spin-dependent $Q^2$ evolution equations [2].
A brute-force method was studied for solving
spin-independent evolution equations and for those with
parton-recombination effects [3].
The same method is applied to the spin-dependent case.
We divide variables into small steps
($N_x$ steps in the Bjorken variable $\ln x$ and
 $N_t$ steps in the variable $t$)
and calculate differentiation and integration
in the evolution equations. They are simply defined by
$df(x)/dx=[f(x_{m+1})-f(x_m)]/\Delta x_m$ and
$\int dx f(x)=\sum_{m=1}^{N_x} \Delta x_m f(x_m)$.
In this way, the integrodifferential equations are solved 
step by step if initial parton distributions are provided.
We also studied a Laguerre polynomial method [4] as an
alternative one in the unpolarized case.
Although the computing time is very short in the Laguerre method,
convergence is not good for ``valence-like" distributions
at small $x$. However, it should be mentioned that 
the Laguerre is still an excellent method in handling 
singlet-quark and gluon distributions.
We decided to apply the brute-force method to the polarized case first
by regarding the accuracy, instead of the computing time, as important.

There are various parametrizations for polarized
parton distributions. Because only available data are
$g_1$ for the proton and deuteron, it is impossible 
at this stage to have accurate information on each
parton polarization. We use one of popular distributions, 
the Gehrmann-Stirling set A [5], which are given at $Q^2=$4 GeV$^2$.
We study evolution of nonsinglet, singlet, and gluon distributions
to $Q^2$=200 GeV$^2$ with $N_f$=4 and $\Lambda_{\overline{MS}\, }$=231 MeV.
For example, results of the singlet-quark evolution 
are shown in Figs. 1 and 2.
Numerical results depend on the step numbers, $N_t$ and $N_x$.
In Fig. 1, $N_t$ is varied from 20 to 1000 steps with fixed $N_x$=1000.
As it is obvious from the figure, the evolved distributions are almost
the same. It means that merely $N_t$=50 steps are enough for getting
accurate evolution. This conclusion is expected because
the scaling violation is a small logarithmic effect.
Next, $N_t$ is fixed at 200 steps and $N_x$ is varied from
100 to 4000 steps in Fig. 2.
From this figure, we find that several hundred $x$ steps are necessary
for obtaining good accuracy.
Therefore, if we choose the parameters $N_t$=200 and $N_x$=1000,
the evolution results are good enough.
We also analyzed evolution of nonsinglet and gluon distributions.
The obtained results show similar accuracy.
It should be mentioned that 
our analysis are still in progress, so that presented numerical
results should be considered preliminary at this stage.

\vspace{-0.8cm}
\noindent
\begin{figure}[h]
\parbox[b]{0.46\textwidth}{
   \begin{center}
       \epsfig{file=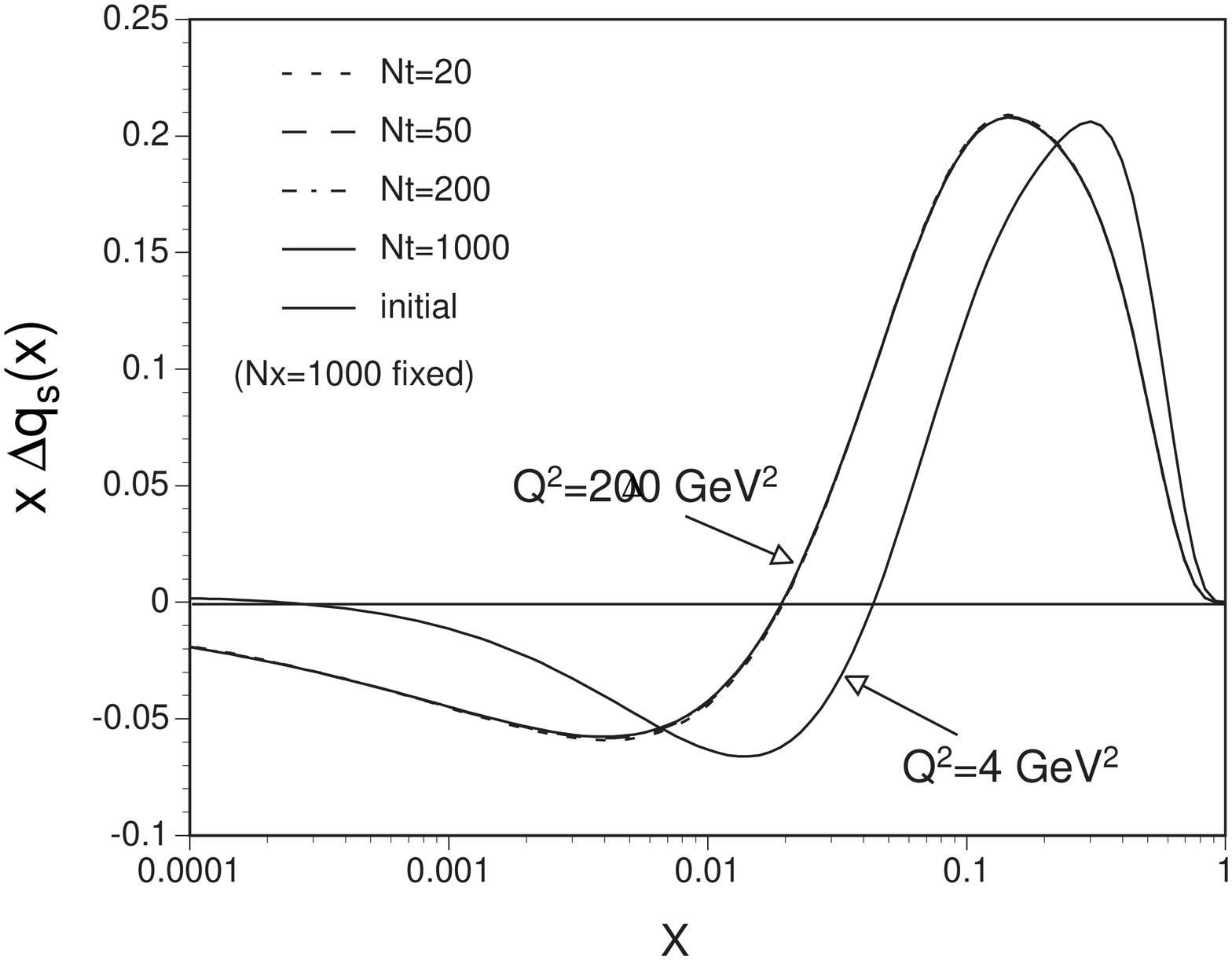,width=6.0cm}
   \end{center}
   \vspace{-0.8cm}
       \caption{\footnotesize
          $N_t$ dependence in $x \Delta q_s$ evolution.}
       \label{fig:fig1}
}\hfill
\parbox[b]{0.46\textwidth}{
   \begin{center}
   \epsfig{file=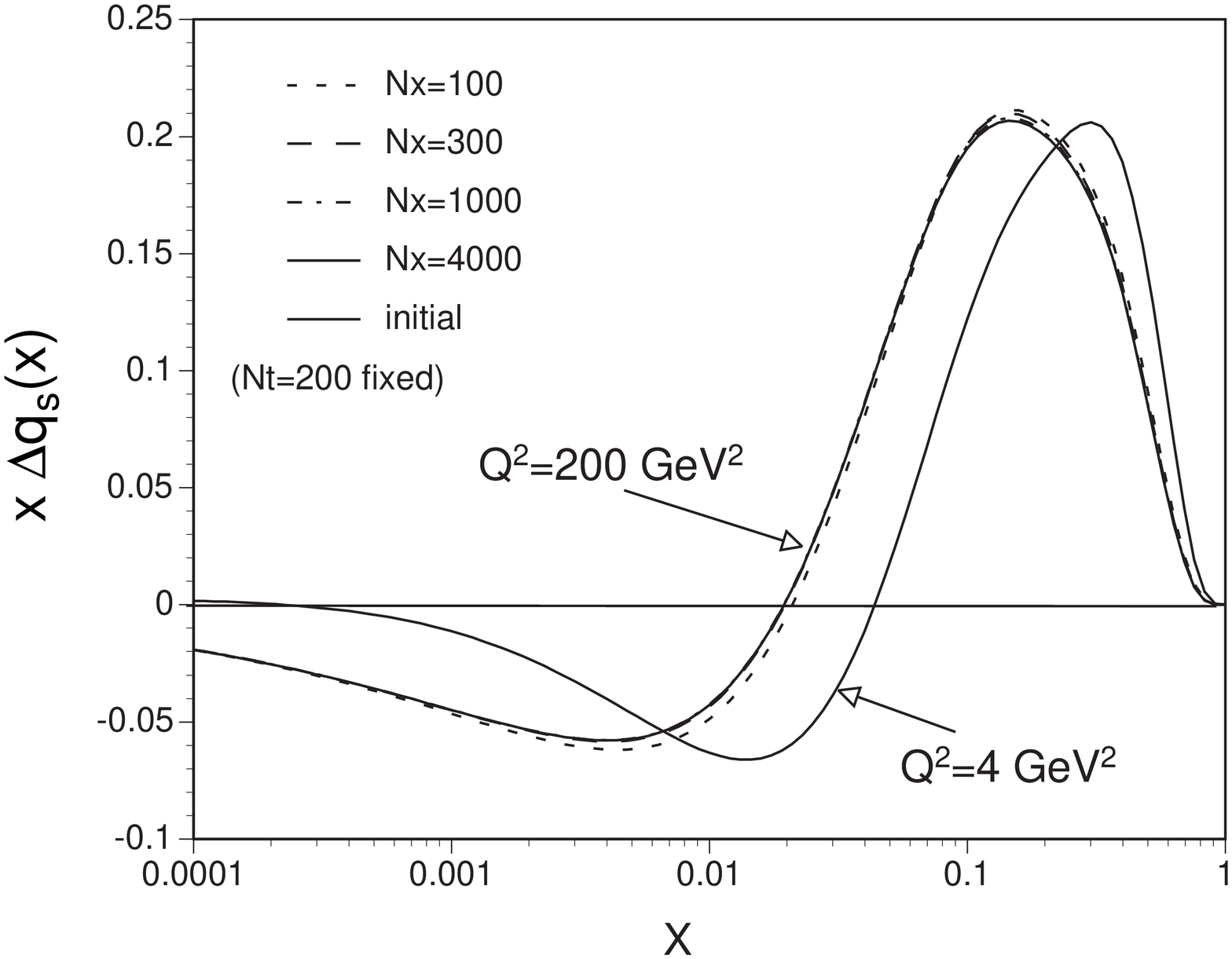,width=6.0cm}
   \end{center}
   \vspace{-0.8cm}
       \caption{\footnotesize 
          $N_x$ dependence in $x \Delta q_s$ evolution.}
       \label{fig:fig2}
}
\end{figure}

Experimental information on the polarized distributions is, for example,
given by the structure function $g_1$. 
In the renormalization scheme $\overline{MS}$,
the $g_1$ should be calculated by the convolution
of polarized parton distributions with coefficient functions:
$$
g_1(x,Q^2)=\sum_{i} e_i^2 g_{1,i}^+(x,Q^2) \ \ \ ,
\eqno{(2a)}
$$
$$
g_{1, i}^+ (x,Q^2)=\int_x^1 {dy \over y}  
\Delta C_1^q \left({{x \over y},{\alpha }_{s}}\right) 
 \Delta q_{i}^+ (y,{Q}^{2})
+\int_{x}^{1}{dy \over y} 
\Delta C_1^g \left({{x \over y},{\alpha }_{s}}\right)
 \Delta g (y,{Q}^{2})
\ ,
\eqno{(2b)}
$$
where $\Delta q_i^+= \Delta q_i + \Delta \bar q_i$.

\begin{wrapfigure}{r}{0.46\textwidth}
  \vspace{-0.3cm}
   \begin{center}
   \epsfig{file=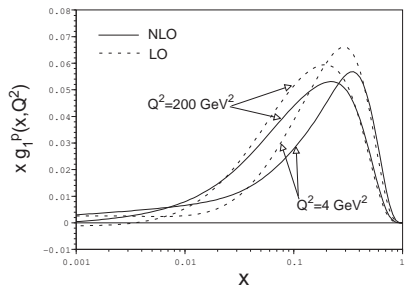,width=6.0cm}
   \end{center}
   \vspace{-0.9cm}
       \caption{\footnotesize 
          $Q^2$ evolution of $xg_1$.}
       \label{fig:fig3}
\end{wrapfigure}
We show results of the $g_1$ evolution in Fig. 3.
The initial distributions are given again by the GS set A,
and $g_1$ at $Q^2$=4 GeV$^2$ is calculated by Eqs. (2a) and (2b).
The $g_1$ at $Q^2$=200 GeV$^2$ is calculated 
first by evolving the initial distributions, then
by taking into account the coefficient functions by
Eqs. (2a) and (2b).
We fix the parameters $N_t$=200 and $N_x$=1000
and show the evolved structure functions in 
the leading order (LO) and in the NLO.
Note that the GS distributions, which are obtained in the NLO
analysis, are also used in the LO input distributions at $Q^2$=4 GeV$^2$.
The NLO contributions are conspicuous at small $x$.
Because the gluon polarization should
be extracted from the scaling violation at small $x$, 
our NLO studies are important.

Our numerical analysis are still in progress, particularly in comparison
with scaling violation data of $g_1$.
However, the results indicate that accurate solution is
obtained in the region $10^{-4}<x<0.8$
by taking more than two-hundred $t$ steps
and more than one-thousand $x$ steps.

\vspace{-0.2cm}
\begin{center}
{\bf Acknowledgments} \\
\end{center}
\vspace{-0.35cm}

SK thanks RIKEN for its financial support 
for participating in this conference.
This research was partly supported by the Grant-in-Aid for
Scientific Research from the Japanese Ministry of Education,
Science, and Culture under the contract number 06640406.

\vspace{0.2cm}
\noindent
{* Email: 96sm18, kumanos, 96td25@cc.saga-u.ac.jp;}  \\

\vspace{-0.55cm}
{http://www.cc.saga-u.ac.jp/saga-u/riko/physics/quantum1/structure.html.} \\

\small

\vspace{-0.4cm}

\begin{center}
{\bf References} \\
\end{center}

\vspace{-0.75cm}
\begin{description}
\item{[1]} R. Mertig and W. L. van Neerven, Z. Phys. C70 (1996) 637;
           W. Vogelsang, Phys. Rev. D54 (1996) 2023; RAL-TR-96-020.

\vspace{-0.30cm}
\item{[2]}
M. Hirai, S. Kumano, and M. Miyama, research in progress.

\vspace{-0.30cm}
\item{[3]}
M. Miyama and S. Kumano, 
              Comput. Phys. Commun. 94 (1996) 185.

\vspace{-0.30cm}
\item{[4]}
R.Kobayashi, M.Konuma, and S.Kumano, 
              Comput. Phys. Commun. 86 (1995) 264. 

\vspace{-0.30cm}
\item{[5]} T. Gehrmann and W. J. Stirling, Phys. Rev. D53 (1996) 6100.

\end{description}

\end{document}